\documentclass[aps,pre,10pt,twocolumn,amsmath,citeautoscript,groupedaddress,superscriptaddress]{revtex4-2}

\usepackage[colorlinks,urlcolor=blue,citecolor=blue]{hyperref}
\usepackage{color}
\usepackage[export]{adjustbox}
\usepackage{graphicx}
\usepackage{amsmath,amssymb}
\usepackage[per-mode=symbol]{siunitx}
\DeclareSIUnit\angstrom{\text{\AA}}

\begin{document}

\title{Trajectory sampling and finite-size effects in first-principles\\ stopping power calculations}

\author{Alina Kononov}
\affiliation{Center for Computing Research, Sandia National Laboratories, Albuquerque NM, USA}
\author{Thomas Hentschel}
\affiliation{School of Applied \& Engineering Physics, Cornell University, Ithaca NY, USA}
\author{Stephanie B. Hansen}
\affiliation{Pulsed Power Sciences Center, Sandia National Laboratories, Albuquerque NM, USA}
\author{Andrew D. Baczewski}
\affiliation{Center for Computing Research, Sandia National Laboratories, Albuquerque NM, USA}

\date{\today}

\begin{abstract}
Real-time time-dependent density functional theory (TDDFT) is presently the most accurate available method for computing electronic stopping powers from first principles.
However, obtaining application-relevant results often involves either costly averages over multiple calculations or ad hoc selection of a representative ion trajectory.
We consider a broadly applicable, quantitative metric for evaluating and optimizing trajectories in this context.
This methodology enables rigorous analysis of the failure modes of various common trajectory choices in crystalline materials.
Although randomly selecting trajectories is common practice in stopping power calculations in solids, we show that nearly 30\% of random trajectories in an FCC aluminium crystal will not representatively sample the material over the time and length scales feasibly simulated with TDDFT, and unrepresentative choices incur errors of up to 60\%.
We also show that finite-size effects depend on ion trajectory via ``ouroboros'' effects beyond the prevailing plasmon-based interpretation, and we propose a cost-reducing scheme to obtain converged results even when expensive core-electron contributions preclude large supercells.
This work helps to mitigate poorly controlled approximations in first-principles stopping power calculations, allowing 1\,--\,2 order of magnitude cost reductions for obtaining representatively averaged and converged results.
\end{abstract}

\maketitle

\section{Introduction}
\label{sec:intro}

High-performance computing has revolutionized materials science, enabling prediction, design, and unprecedented understanding of materials properties to complement and accelerate experimental efforts.
Modeling dynamic, nonlinear responses to stimuli such as laser and particle irradiation falls among the most computationally demanding types of materials simulations, requiring real-time evolution of extended systems containing hundreds of atoms and thousands of quantum-mechanical electrons \cite{kang:2019,shepard:2021,kononov:2022}.
For materials in extreme conditions, high temperatures result in orders-of-magnitude increases in the number of partially occupied electronic orbitals, either requiring additional approximations \cite{white:2018,white:2020,white:2022} or further escalating computational resource requirements to millions of CPU-hours or more per calculation \cite{hentschel:2023}.
In this context, deliberate design of simulations is crucial for maximizing insight while maintaining feasible computational costs.

Here, we focus on calculations of electronic stopping power, the rate at which a moving particle loses energy to electrons.
This fundamental quantity is critically important to diverse fields.
For example, radiation therapy relies on stopping powers to predict particle ranges and precisely target tumors \cite{parodi:2018}.
Stopping powers also underlie radiation damage to materials in space and nuclear energy applications \cite{duzellier:2005, allen:2010}.
In materials imaging and processing techniques, energy deposition by focused ion beams relates to electron emission, sample damage, and defect engineering \cite{hlawacek:2014,li:2017}.
Finally, achieving ignition in fusion energy research relies on fusion products redepositing their kinetic energy into the fuel \cite{zylstra:2019}.

First-principles simulations using real-time time-dependent density functional theory (TDDFT) can offer accurate predictions of electronic stopping powers and insights into underlying physical processes \cite{schleife:2015,lim:2016,quashie:2016,yao:2019,kononov:2020,kononov:2021,shepard:2023}, often in more detail than possible experimentally.
However, computing average stopping powers that are comparable to experimentally observable and practically relevant values can pose a challenge.
While an individual TDDFT calculation simulates a single projectile traversing a specific path, stopping power experiments measure energy loss distributions for finite-width ion beams, often incident on polycrystalline or disordered samples \cite{moro:2016,roth:2017,malko:2022exp}.
Moreover, applications either also employ finite-width ion beams (e.g., materials imaging and processing) or involve randomly oriented radiation (e.g., equipment in space and fusion fuel).
Therefore, sensitivities to the projectile's trajectory can limit the utility of TDDFT stopping power predictions.
Such sensitivities often occur for projectile velocities beyond the Bragg peak and mainly affect contributions from core electrons, or more generally, spatially non-uniform electronic orbitals \cite{schleife:2015,ullah:2018,yao:2019}.

\begin{figure}
    \centering
    \includegraphics[trim={0 0.08in 0.2in 0.2in}, clip]{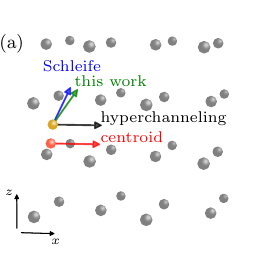}\hfill
    \includegraphics[trim={0 0.08in 0.05in 0.2in}, clip]{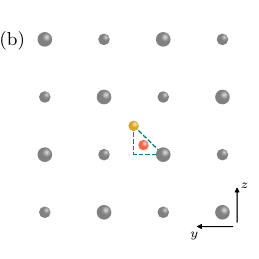}
    \caption{Trajectories through FCC aluminum discussed in this work.
    Red and yellow balls represent the projectile's initial position, and arrows indicate the direction of motion.
    The hyperchanneling and centroid trajectories are parallel to a lattice vector, unlike the off-channeling trajectories considered by Schleife et al.\ \cite{schleife:2015} and in this work.
    In (b), the dashed triangle indicates the symmetry-irreducible cross-section of the lattice.
    The red ball lies at the centroid of this triangle.
    }
    \label{fig:traj_diagram}
\end{figure}

Meaningful average stopping powers can still be obtained from first principles by averaging \cite{yost:2016,yost:2019,yao:2019,li:2021,white:2022} or carefully integrating \cite{maliyov:2018, maliyov:2020} the results of several TDDFT calculations using distinct projectile trajectories.
The significant computational cost of this approach makes it tempting to select a single trajectory presumed to be representative of an ensemble average.
For solids, one possible choice is the centroid trajectory, wherein the projectile travels along a crystallographic direction with a path given by the geometric centroid of a symmetry-irreducible cross-section of the crystal structure \cite{ojanpera:2014,yost:2016} (see Fig.\ \ref{fig:traj_diagram}).
This method appears adequate when core electron contributions are small, but becomes inaccurate for fast projectiles and high-Z targets \cite{maliyov:2020}.
Alternatively, a randomly chosen trajectory can achieve good agreement with empirical data even when core-electron contributions are important \cite{schleife:2015}.
In this case, trajectory choice may constitute an uncontrolled approximation, thus necessitating quantitative methods for assessing any given choice.

Recently, Gu et al.\ developed an innovative pre-sampling approach \cite{gu:2020} that averages results from several short trajectories carefully selected such that, in aggregate, they representatively sample a disordered system.
Here, we present a complementary method that uses a quantitative metric to guide \emph{a priori} selection of a single, representative projectile trajectory for first-principles calculations of electronic stopping power.
Using proton stopping in aluminum as an exemplar, we demonstrate the utility of such an approach even in a crystalline material, as we find that achieving agreement with empirical data requires a high-quality trajectory despite wide-spread presumptions that a random choice suffices.
While both methods can reduce computational costs associated with averaging over multiple long trajectories and help obtain reliable results even in the absence of experimental data to validate against, we expect that our approach will be more efficient than the one proposed by Gu et al.\ \cite{gu:2020} in the case of heavy ions that experience relatively long transient behavior \cite{lee:2020} before entering the steady-state stopping regime.

In addition to trajectory-dependent core-electron contributions, finite-size effects limit the accuracy of first-principles stopping power calculations, particularly for fast projectiles with velocities above the Bragg peak.
Typically, TDDFT calculations are expected to underestimate stopping powers by neglecting long-wavelength plasmonic excitations that a finite periodic supercell cannot support \cite{correa:2018}.
We examine variations in computed stopping powers for different supercell sizes and projectile trajectories and reveal significant departures from this model of finite-size errors.
Finally, we extend our trajectory optimization framework with a second quantitative metric that enables \emph{a priori} selection of trajectories that minimize finite-size effects.

Together, the two trajectory metrics developed in this work allow deeper understanding and new opportunities for mitigation of previously poorly controlled approximations in TDDFT simulations of energetic particles traversing matter.
This contribution advances the accuracy and efficiency of first-principles stopping power calculations with wide-ranging implications for computational studies relating to radiation therapy, materials in extreme conditions, ion-beam imaging and patterning techniques, and self-heating of fusion fuel.

\section{Results}
\label{sec:results}

\subsection{Sampling close collisions}
\label{sec:traj}

Obtaining an application-relevant result from a single TDDFT stopping calculation requires choosing a trajectory along which the projectile experiences an environment that quantitatively resembles those likely to occur in that application.
In particular, close collisions between the projectile and host nuclei can involve (semi)core electron excitations that introduce sharp features in the stopping forces and can dominate the average stopping power \cite{schleife:2015}.
Thus, representatively sampling close collisions is essential for computing accurate stopping powers, and we expect that the distance between the projectile and the nearest host nuclei at all points along the trajectory provides a compact description of the environment that determines the average stopping power.

This notion previously inspired a method for assessing the quality of different trajectories in disordered systems \cite{gu:2020}.
In what follows, we present a quantitative metric that enables rigorous comparisons among different trajectories and their attendant stopping powers.
Furthermore, we will show that such an approach is critical for selecting trajectories even in crystalline materials because not all randomly-oriented off-channeling trajectories are equally representative of an average environment over the course of a typical few-fs simulation time.
Unrepresentative finite-length off-channeling trajectories lead to poor estimates of average stopping power and could skew averages based on naive trajectory sampling.

To evaluate the quality of a given trajectory, we first calculate the distribution $P_{\mathrm{traj}}(\delta_{NN})$ of nearest-neighbor distances $\delta_{NN}$ between the host nuclei and the projectile along the trajectory, excluding data during the first \SI{4}{\angstrom} of the projectile's path since the ultimate stopping power extraction will ignore this early transient regime (see \mbox{Sec.\ \ref{sec:sm:extraction}} of the supplementary information (SI)).
This value is specific to a proton projectile, and higher Z projectiles would have longer transient regimes \cite{lee:2020}, further motivating optimization of a single representative trajectory.
Computing $P_{\mathrm{traj}}(\delta_{NN})$ relies only on the geometric specification of the supercell and does not require expensive TDDFT simulations.

We then compare $P_{\mathrm{traj}}(\delta_{NN})$ to an ideal distribution $P_{\mathrm{ideal}}(\delta_{NN})$ generated by calculating distances to nearest host nuclei from randomly sampled points within the supercell.
This choice of $P_{\mathrm{ideal}}(\delta_{NN})$ represents the distribution experienced by randomly oriented radiation or a focused ion beam interacting with randomly oriented grains within a polycrystalline sample.
Earlier work by Gu et al.\ instead generated a reference distribution by sampling along a \SI{500}{\micro\meter}-long trajectory \cite{gu:2020}, a method that should lead to the same distribution provided that the selected reference trajectory representatively samples the entire supercell.
Other choices may be more suitable depending on the specific application, e.g.\ a focused ion beam aligned with a lattice vector of a single crystal.
Section \ref{sec:sm:dists} in the SI discusses numerical sampling of $P_\mathrm{ideal}$ and $P_\mathrm{traj}$.

The Hellinger distance $D_H$ between the two nearest-neighbor distributions, given by
\begin{equation}
    D_H^2 = 1 - \int_0^\infty \sqrt{P_{\mathrm{traj}}(\delta_{NN}) P_{\mathrm{ideal}}(\delta_{NN})} \; d\delta_{NN},
    \label{eq:dh}
\end{equation}
provides a quantitative measure of how well a trajectory samples the simulation cell.
Notably, $D_H$ is bounded between $0$ and $1$, with $D_H=1$ achieved when the two distributions have no overlap and $D_H=0$ achieved only when the two distributions are identical.
Minimizing $D_H$ will enable selection of optimally representative trajectories.
We chose the Hellinger distance over other possible ways to quantify the similarity of two distributions because it satisfies the mathematical properties of a metric.
In particular, $D_H$ is symmetric and obeys the triangle inequality, allowing sensible comparisons of two different trajectories by replacing $P_{\mathrm{ideal}}(\delta_{NN})$ in Eq.\ \eqref{eq:dh} with the nearest-neighbor distribution of the second trajectory.
We compare the $D_H$ metric used in this work with the overlap index used by Gu et al.\ \cite{gu:2020} in Section \ref{sec:sm:gu_comparison} of the SI.

As the proton travels, $P_{\mathrm{traj}}(\delta_{NN})$ evolves and $D_H$ becomes a function of the total distance traversed by the projectile.
All but a few pathological random trajectories will asymptotically approach $D_H=0$ as the total distance traversed tends to infinity, but the impact of finite-size effects (e.g., spurious interactions between the projectile and its wake) will also grow with the total distance.
Thus, a ``good'' trajectory should achieve a small $D_H$ after the projectile travels a relatively short distance so that a single, reasonably short TDDFT calculation suffices to obtain an accurate estimate of the average stopping power.
In the following, we investigate the behavior of $D_H$ for a range of trajectories and the implications for stopping power results.

\begin{figure}
    \centering
    \includegraphics{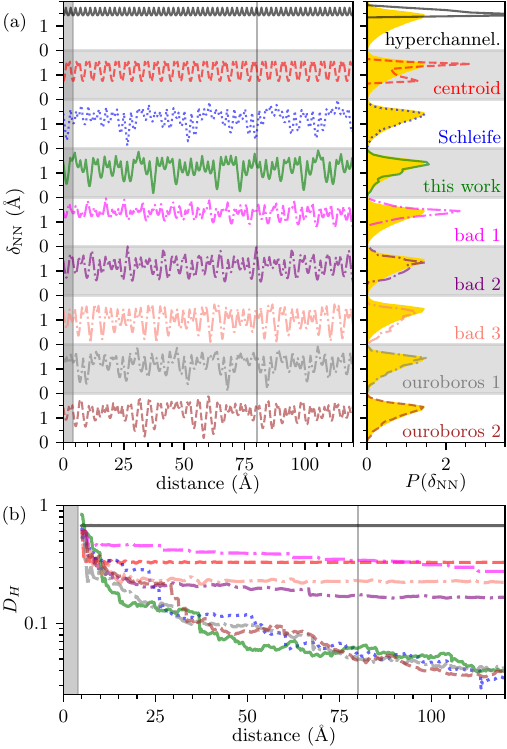}
    \caption{
    (a) Nearest-neighbor distances (left) and distributions (right) experienced by a proton traversing ambient aluminum along various trajectories discussed in this work.
    The nearest-neighbor distribution after the projectile travels \SI{80}{\angstrom} along each trajectory is compared to the ideal distribution sampled from random points within the supercell (yellow).
    Individual panels show data for the hyperchanneling trajectory (black), the centroid trajectory (red), the off-channeling trajectory used in \mbox{Ref.\ \onlinecite{schleife:2015} (blue)}, the ``good" off-channeling trajectory used in this work (green), three ``bad" off-channeling trajectories that do not achieve a representative near-neighbor distribution, and two other off-channeling trajectories discussed in the context of ouroboros effects in Sec.\ \ref{sec:finitesize}.
    (b) Hellinger distance (see Eq.\ \eqref{eq:dh}) from the ideal distribution achieved by each trajectory as a function of total distance traveled by the projectile.
    The gray area indicates the transient regime ignored in the analysis.
    }
    \label{fig:traj_metric}
\end{figure}

In Fig.\ \ref{fig:traj_metric}a we consider the case of FCC aluminum and compare the nearest-neighbor distances and corresponding distributions $P_{\mathrm{traj}}(\delta_{NN})$ occurring along several trajectories, including those illustrated in Fig.\ \ref{fig:traj_diagram}.
Both the hyperchanneling and centroid trajectories lie parallel to a lattice vector and thus sample periodic environments, producing nearly static nearest-neighbor distributions and asymptotically constant $D_H$ values of about 0.67 and 0.33, respectively (see Fig.\ \ref{fig:traj_metric}b).
Of course, $D_H$ remains relatively large for the hyperchanneling trajectory because its nearest-neighbor distribution is severely skewed toward large $\delta_{NN}$.

\begin{figure}
    \centering
    \includegraphics[valign=t]{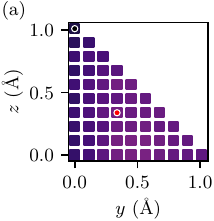}
    \hfill
    \includegraphics[valign=t]{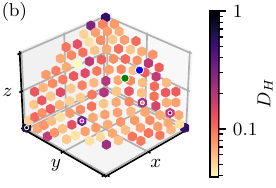}
    \caption{The trajectory metric $D_H$ after a proton travels \SI{80}{\angstrom} along different trajectories.
    In (a), the proton moves in the $x$ direction and its initial position is varied within a symmetry-irreducible cross-section of the lattice (see Fig.\ \ref{fig:traj_diagram}b).
    Black and red points correspond to the hyperchanneling and centroid trajectories, respectively.
    In (b), the proton begins at a high-symmetry point within the lattice (the yellow ball in Fig.\ \ref{fig:traj_diagram}) and the direction of its velocity is varied over an octant of the unit sphere.
    The $x$, $y$, and $z$ directions correspond to channeling trajectories.
    Black, blue, and green points indicate the hyperchanneling trajectory, the trajectory used by Schleife et al.\ \cite{schleife:2015}, and the trajectory used in this work, respectively.
    Pink and purple points with white outlines indicate ``bad" trajectories tested in this work.
    }
    \label{fig:traj_scan}
\end{figure}

While the centroid trajectory significantly improves $D_H$ over the hyperchanneling trajectory, its nearest-neighbor distribution is bimodal, oversampling near points of closest and furthest approach at $\delta_{NN}\approx 0.75$ and \SI{1.5}{\angstrom} while lacking close collisions with $\delta_{NN}< 0.5$.
Failure to capture the ideal nearest-neighbor distribution explains the poor performance reported for the centroid trajectory in regimes where core electron excitations contribute significantly to electronic stopping power \cite{maliyov:2020}.
We also find that other channeling trajectories with different impact parameters are similarly restricted to $D_H>0.3$ (see Fig.\ \ref{fig:traj_scan}a) and therefore do not representatively sample the supercell.

\begin{figure}
    \centering
    \includegraphics{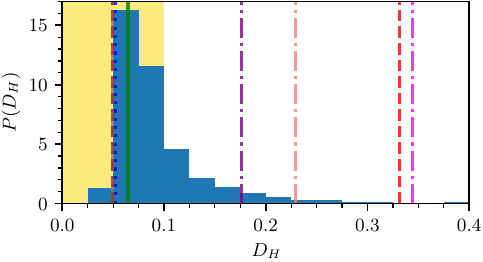}
    \caption{Distribution of $D_H$ values achieved by 1440 random trajectories with initial positions uniformly sampled within the FCC aluminum supercell and directions of motion uniformly sampled over the unit sphere.
    Vertical lines indicate $D_H$ values achieved by the trajectories analyzed in Fig.\ \ref{fig:traj_metric}.
    In all cases, $D_H$ is taken after the projectile travels \SI{80}{\angstrom}.
    Yellow shading indicates $D_H<0.1$, the target identified for representative sampling of the supercell in this work.
    }
    \label{fig:traj_random}
\end{figure}

In contrast, both the off-channeling trajectory considered by Schleife et al.\ \cite{schleife:2015} and the other ``good" off-channeling trajectory identified in this work approximate the ideal distribution more closely and continue to reduce $D_H$ as the proton travels farther, reaching much lower $D_H$ values of 0.05\,--\,0.06 by \SI{80}{\angstrom}.
Even when the projectile starts at a high-symmetry point within the crystal structure, the vast majority of possible directions of motion achieve $D_H < 0.3$ by the time the projectile travels \SI{80}{\angstrom} (see Fig.\ \ref{fig:traj_scan}b).
Some but not all exceptions lie at channeling directions.
For a finite simulation length, $D_H$ can be very sensitive to trajectory direction, with small changes in the trajectory angle sometimes leading to order-of-magnitude changes in $D_H$.
Among fully random trajectories where the projectile's initial position and direction of motion are both uniformly sampled, only 1.3\% perform as poorly as channeling trajectories with $D_H$ still exceeding 0.3 after the proton traverses \SI{80}{\angstrom} (see Fig.\ \ref{fig:traj_random}).

To verify the utility of our trajectory metric and deduce a threshold $D_H$ value for accurate stopping power predictions, we also consider three ``bad" random trajectories that either undersample (bad 1 and bad 2) or oversample (bad 3) close collisions with $\delta_\mathrm{NN}<0.5$ (see Fig.\ \ref{fig:traj_metric}a).
Unlike channeling trajectories, $D_H$ continues to decrease along these bad off-channeling trajectories, but much more slowly than the good off-channeling trajectories, only achieving $D_H=0.17$\,--\,0.35 by the time the proton travels \SI{80}{\angstrom}.

\begin{figure}
    \centering
    \includegraphics[width=\columnwidth]{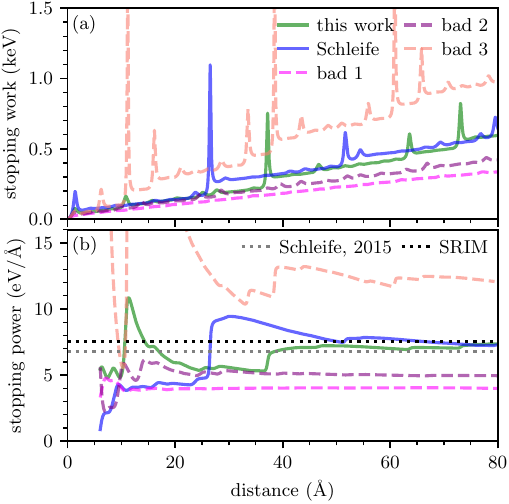}
    \caption{
    Instantaneous (a) stopping work and (b) extracted stopping power as a proton with 4 atomic units of velocity traverses aluminum along the off-channeling trajectory used in Ref.\ \onlinecite{schleife:2015} (blue), a different ``good" off-channeling trajectory used throughout this work (green), and three ``bad" off-channeling trajectories (dashed).
    Results predicted by an earlier TDDFT study \cite{schleife:2015} and the SRIM empirical model \cite{ziegler:2010} are shown as gray and black dotted lines, respectively.
    }
    \label{fig:trajS}
\end{figure}

Finally, we perform TDDFT stopping power calculations as described in Sec.\ \ref{sec:methods} for the off-channeling trajectories examined above.
Although the two good trajectories exhibit differing dynamical behavior, the average stopping powers extracted after the proton travels \SI{80}{\angstrom} agree within 1\% and reproduce empirical data from the SRIM database \cite{ziegler:2010} within 3\% (see Fig.\ \ref{fig:trajS}).
Meanwhile, stopping powers computed using the bad trajectories deviate from empirical data by up to 60\%.
As expected, trajectories that undersample close collisions (bad 1 and bad 2) underestimate stopping power while the trajectory that oversamples close collisions (bad 3) overestimates stopping power.
Based on these findings, we propose that $D_H\lesssim 0.1$ suffices for representative sampling of the nearest-neighbor distribution in these stopping power calculations.
Notably, 27\% of random trajectories still exceed this threshold after the projectile travels \SI{80}{\angstrom} (see \mbox{Fig.\ \ref{fig:traj_random}}), highlighting the importance of careful trajectory selection for accurate and efficient stopping power predictions.

\subsection{Mitigating finite-size effects}
\label{sec:finitesize}

\begin{figure}
    \centering
    \includegraphics{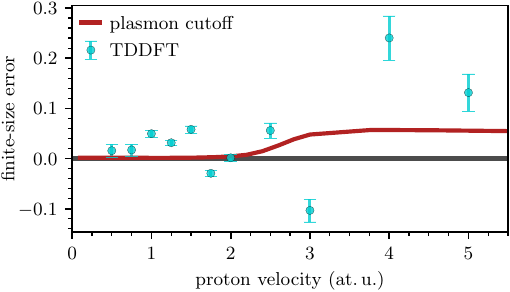}
    \caption{
    Fractional finite-size errors estimated as the relative difference between free-electron stopping powers computed with (12.15\,\AA)$^3$ and (16.2\,\AA)$^3$ cubic supercells, $(S(16.2) - S(12.15))/S(16.2)$.
    TDDFT results are compared to dielectric-based predictions using corresponding plasmon wavelength cutoffs (see Eq.\ \eqref{eq:lr} and accompanying text).
    Error bars estimate uncertainties in the TDDFT data arising from variations in extracted average stopping powers as a function of total simulation time.
    }
    \label{fig:3efs}
\end{figure}

Finite periodic supercells limit the wavelength of plasmonic excitations that a plane-wave TDDFT calculation can capture, often leading to underestimated electronic stopping powers at high projectile velocities \cite{correa:2018}.
Finite-size errors caused by this plasmon cutoff can be estimated from linear response theory \cite{lindhard:1964},
which describes the stopping power in terms of the frequency and wave-vector dependent dielectric function $\epsilon(k, \omega)$:
\begin{equation}
    S(v) = \frac{2Z^2}{\pi v^2} \int_0^\infty \frac{dk}{k} \int_0^{kv} d\omega \,\omega\, \mathrm{Im} \left[ \frac{-1}{\epsilon(k, \omega)}\right],
    \label{eq:lr}
\end{equation}
where $Z=1$ and $v$ are the projectile charge and velocity, respectively.
Here, we employ the Mermin model dielectric function \cite{mermin:1970} with a constant electron-ion collision frequency of 0.1\,at.\,u.
We estimate the contribution of the long-wavelength plasmons that the TDDFT calculations neglect by imposing an upper limit of $k_\mathrm{cut} = 2\pi/L$ for the $k$ integral in Eq.\ \eqref{eq:lr}, where $L$ is the length of the cubic supercell, and evaluating the integrals numerically as recently described in Ref.\ \onlinecite{hentschel:2023}.
This portion of the total stopping power becomes significant when the integration limits contain the plasmon pole, i.e., when $k_\mathrm{cut}v \gtrsim \omega_p$, where $\omega_p \approx$ \SI{16}{\electronvolt} is the aluminum plasma frequency.
Indeed, for $L=$\SI{12.15}{\angstrom} the linear response formalism predicts sizeable finite-size errors for velocities above $\omega_p/k_\mathrm{cut} \approx 2$\,at.\,u.\ (see Fig.\ \ref{fig:3efs}).

However, we find that differences between stopping powers computed using TDDFT with different size supercells only loosely correlate with finite-size errors estimated from linear response.
As shown in Fig.\ \ref{fig:3efs}, the relative difference between TDDFT results calculated using (12.15\,\AA)$^3$ and (16.2\,\AA)$^3$ cubic supercells often significantly exceeds the values predicted using the dielectric model.
Moreover, for some proton velocities, the smaller supercell actually produces greater stopping powers than the larger supercell, a result entirely inconsistent with the plasmon-based interpretation of finite-size effects.

\begin{figure}
    \centering
    \includegraphics{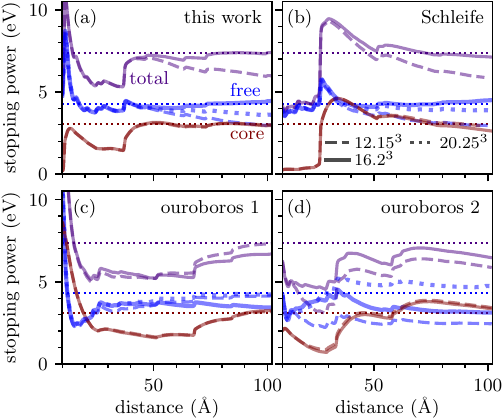}
    \caption{
    Average stopping powers as a proton with 4 atomic units of velocity traverses aluminum along four different off-channeling trajectories that achieve $D_H<0.1$ by the time the proton travels \SI{80}{\angstrom}.
    Results are compared for (a) the trajectory used throughout this work, (b) the trajectory used in Ref.\ \onlinecite{schleife:2015}, and (c-d) two other off-channeling trajectories.
    Total stopping powers (purple) are decomposed into contributions from free (blue) and core (red) electrons for the 256-atom, (\SI{16.2}{\angstrom})$^3$ supercell used throughout this work (solid) and a smaller, 108-atom, (\SI{12.15}{\angstrom})$^3$ supercell (dashed).
    Free-electron results for a larger, 500-atom, (\SI{20.25}{\angstrom})$^3$ supercell are also shown (dotted).
    Horizontal lines indicate converged values for the first trajectory with a (\SI{16.2}{\angstrom})$^3$ supercell.
    }
    \label{fig:fs}
\end{figure}

As it turns out, the choice of trajectory not only affects fidelity of close collision sampling, but also influences finite-size errors, an effect not captured by the plasmonic model.
In Fig.\ \ref{fig:fs}, we compare stopping powers computed using three different size supercells and four different proton trajectories that each achieve comparably small $D_H$ values (see Fig.\ \ref{fig:traj_metric}).
Contributions from free (conduction) and core ($2s$ and $2p$) electrons were isolated through the use of different pseudopotentials (see \mbox{Sec.\ \ref{sec:methods}} for more details).
Close agreement of the converged core-electron stopping powers verifies that each trajectory adequately samples close collisions with aluminum ions.

While the core-electron contribution shows little variation with supercell size, the free-electron stopping powers are quite sensitive to finite-size effects.
Similar to the negative finite-size errors appearing in Fig.\ \ref{fig:3efs}, the computed free-electron stopping powers do not always grow monotonically with increasing supercell size: for the trajectories used throughout this work and \mbox{Ref.\ \onlinecite{schleife:2015}}, the largest, (\SI{20.25}{\angstrom})$^3$ supercell leads to somewhat smaller free-electron stopping powers than the intermediate, (\SI{16.2}{\angstrom})$^3$ supercell (see Fig.\ \ref{fig:fs}a and b).
Meanwhile, in \mbox{Fig.\ \ref{fig:fs}c}, the smallest, (\SI{12.15}{\angstrom})$^3$ supercell produces a greater free-electron stopping power than the (\SI{16.2}{\angstrom})$^3$ supercell.
The magnitude of discrepancies between results computed with different size supercells also depends on the projectile trajectory.

We attribute the surprising trajectory-dependent and nonmonotic behavior of finite-size effects in TDDFT stopping power calculations to artificial interactions with previously excited electrons.
In particular, if the projectile passes near its earlier path after re-entering a periodic supercell, then it interacts with an excited electron density rather than pristine material, distorting stopping power results.
Such ``ouroboros" effects \footnote{We suggest ``Pac-Man'' effects as an alternative to ``ouroborous'' effects for any ophidiophobic readers.} are especially severe for channeling trajectories: upon reentry, the projectile traverses the exact same path, at which point the instantaneous stopping power begins to depend on supercell size \cite{schleife:2015}.
Ouroboros effects can also pollute results for off-channeling trajectories, since the projectile has a finite interaction radius that may partially overlap with previously excited regions.
Even if the projectile remains relatively far from previously traversed material, it could still interact with earlier electronic excitations that have propagated into its path.
We distinguish these different types of artificial interactions as static and dynamic ouroborous effects, respectively.

The prevalence of ouroboros effects can be estimated by considering the minimum distance $D_O$ between periodic images of the projectile's path.
This distance is upper-bounded by the simulation cell dimensions and decreases as the projectile crosses periodic boundaries.
In the limit of large $D_O$, the projectile remains far from earlier excited material and ouroboros effects should be minimal.
In a metallic system, localized excitations only involve core atomic orbitals and are well-confined within the Wigner-Seitz radius $r_{\mathrm{WS}}$ of the host nuclei.
Therefore, we expect that ensuring $D_O > r_{\mathrm{WS}}$ (\SI{1.58}{\angstrom} for FCC aluminum) eliminates static ouroboros effects.
Even for the lowest $D_O$ tested, \SI{1.3}{\angstrom} for the smallest supercell in Fig.\ \ref{fig:fs}d, finite-size effects do not influence the core-electron stopping power because the $2s$ and $2p$ electrons are further localized within about \SI{0.5}{\angstrom} of aluminum nuclei.

Dynamic ouroboros effects, on the other hand, are much harder to characterize and avoid.
Given a Fermi velocity of about \SI{20}{\angstrom\per\femto\second} in aluminum, single-particle excitations could traverse the (\SI{16.2}{\angstrom})$^3$, 256-atom supercells used throughout this work in less than \SI{1}{\femto\second}, a time scale comparable to the duration of the TDDFT simulations.
First-principles calculations of plasmon dispersion in aluminum \cite{ramakrishna:2021} suggest similar propagation speeds for collective excitations.
Since an off-channeling projectile must travel around \SI{80}{\angstrom} in order to adequately sample the nearest-neighbor distribution in this material (see Sec.\ \ref{sec:traj}), it crosses the periodic boundaries multiple times, leading to typical $D_O$ distances of around 3\,\AA\ that electronic excitations traverse over an even shorter time scale.

\begin{figure}
    \centering
    \includegraphics{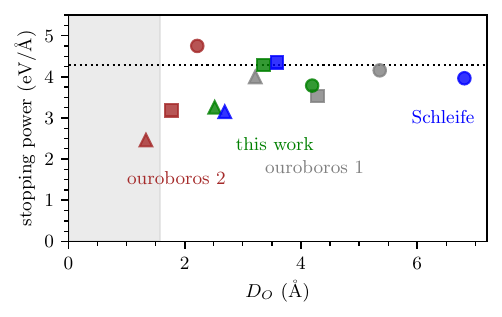}
    \caption{Convergence of free-electron stopping power with increasing $D_O$, the minimum distance between periodic images of the projectile's path, for a proton traversing aluminum with 4 atomic units of velocity.
    Both the stopping powers and $D_O$ values reported were taken after the projectile traveled \SI{80}{\angstrom} through the material.
    Colors indicate proton trajectory, while symbols indicate supercell size: triangles, squares, and circles correspond to results from (\SI{12.15}{\angstrom})$^3$, (\SI{16.2}{\angstrom})$^3$, and (\SI{20.25}{\angstrom})$^3$ supercells, respectively.
    The shaded area indicates $D_O < r_\mathrm{WS}$, and the dotted line indicates stopping power extracted from the simulation parameters used throughout this work.
    }
    \label{fig:DO}
\end{figure}

However, proton projectiles can be expected to induce fairly weak charge perturbations, as evidenced by the relative success of linear response treatments of proton stopping powers \cite{schleife:2015,peralta:2022,hentschel:2023}.
So, dynamic ouroboros effects could be small compared to other sources of error.
Furthermore, alternating artificial interactions with excitations that involve excess or reduced electron density relative to the pristine material could have partially cancelling influences on the average stopping power.
Indeed, the free-electron stopping powers reported in Fig.\ \ref{fig:fs} differ among each other by at most 31\% of the total stopping power, whereas poor sampling of close collisions affected total stopping powers by almost a factor of 2 in Fig.\ \ref{fig:trajS}.
In Fig.\ \ref{fig:DO}, we show that $D_O\gtrsim$ \SI{3}{\angstrom} already achieves acceptable convergence of free-electron stopping powers.

The fact that finite-size effects predominantly influence free-electron contributions to stopping power can be exploited to reduce computational costs associated with first-principles stopping power calculations.
Core contributions typically dominate computational costs because explicitly modeling core electrons dramatically increases the spectral range of the Kohn-Sham Hamiltonian, generally requiring higher plane-wave cutoff energies, smaller time steps, and/or more solver iterations in implicit time-stepping algorithms such as the one used in this work \cite{baczewski:2014}.
Since core contributions are not very sensitive to ouroboros effects or supercell size, they can be efficiently calculated using smaller supercells.
Meanwhile, free-electron contributions can be separately converged with respect to the $D_O$ metric and supercell size while pseudizing core electrons to allow cheaper time evolution.

In this work, applying this scheme to separately compute core and free-electron contributions using (\SI{12.15}{\angstrom})$^3$ and (\SI{16.2}{\angstrom})$^3$ supercells, respectively, would have allowed a seven-fold speedup over using (\SI{16.2}{\angstrom})$^3$ supercells to compute both core and free-electron contributions simultaneously.
In fact, this strategy would have still resulted in a nearly three-fold speedup if the larger, (\SI{20.25}{\angstrom})$^3$ supercell had been used to further reduce finite-size effects in the free-electron contribution.
These savings are relative to the roughly $5\times 10^5$ CPU-hour cost per production calculation in this work, which included core contributions in (\SI{16.2}{\angstrom})$^3$ supercells.

\section{Discussion}

\begin{figure}
\includegraphics[width=\columnwidth]{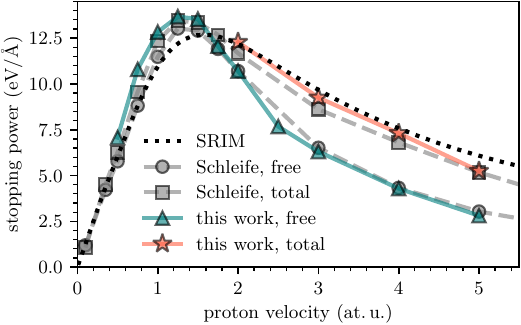}
\caption{
Electronic stopping power of protons in ambient aluminum as predicted by this work, an earlier RT-TDDFT study \cite{schleife:2015}, and the SRIM empirical model \cite{ziegler:2010}.
The free-electron contribution was computed by pseudizing all but the outermost 3 electrons per aluminum atom, while the total stopping power also included excitations of explicitly modeled $2s$ and $2p$ semicore electrons.
}
\label{fig:Scold}
\end{figure}

Fig.\ \ref{fig:Scold} compares our final electronic stopping results as a function of proton velocity to those reported in an earlier TDDFT study \cite{schleife:2015} and the SRIM empirical model \cite{ziegler:2010}.
Overall, we find good quantitative agreement between the two TDDFT datasets, with modest discrepancies arising from a combination of partially cancelling factors.
First, Ref.\ \onlinecite{schleife:2015} verified convergence with respect to plane-wave cutoff energy for a channeling trajectory, but we show in \mbox{Sec.\ \ref{sec:sm:convergence}} of the SI that converging energy transferred during close collisions occurring along off-channeling trajectories requires higher cutoffs.
In particular, we find that the \SI{680}{\electronvolt} cutoff used in Ref.\ \onlinecite{schleife:2015} underestimates high-velocity stopping power by about 5\%.

Another source of discrepancies arises from different pseudopotentials: this work used the PAW method \cite{blochl:1994} within an extension of VASP \cite{baczewski:2016,magyar:2016}, while Ref.\ \onlinecite{schleife:2015} used norm-conserving pseudopotentials \cite{vanderbilt:1985,troullier:1991} within Qb@ll \cite{schleife:2012,draeger:2017}.
In Sec.\ \ref{sec:sm:pps} of the SI, we show that even when all other parameters are fixed or separately converged as appropriate, the PAW and VASP methodology used in this work produces about 10\% lower stopping powers than the harder, norm-conserving pseudopotentials within Qb@ll.
Additionally, a 2.5\% uncertainty in extracted stopping powers arises from variations with respect to the data range included in the stopping power extraction (see Sec.\ \ref{sec:sm:extraction} of the SI).
Further benchmarking of different TDDFT codes, pseudopotentials, basis sets, and other methodological details will be important for reducing uncertainties in first-principles stopping data \cite{malko:2022whitepaper}.

To enable detailed interpretation of discrepancies in computed results and reduce computational costs associated with obtaining representative average stopping powers, we presented a quantitative metric to evaluate the quality of ion trajectories in first-principles electronic stopping power calculations.
The approach resembles the one proposed by Ref.\ \onlinecite{gu:2020} and considers the distribution of nearest-neighbor distances experienced by the ion along its path, which allows scrutiny of how representatively different trajectories sample the close collisions that determine core-electron contributions.
Optimizing trajectories via this metric can help compute stopping powers more accurately and efficiently, particularly for high-Z elements requiring careful sampling of core-electron excitations.
We expect that straightforward extensions of this metric to disordered systems at high temperatures and heterogeneous systems including compounds, alloys, and mixtures will be even more impactful.

We also identified a cost-reducing scheme to systematically characterize finite-size errors.
Analyzing velocity- and trajectory-dependent finite-size effects revealed behavior inconsistent with the prevailing plasmonic model \cite{correa:2018}, which we explain through ``ouroboros" effects wherein the projectile fictitiously interacts with previously excited electrons.
Thus, we propose considering convergence with increasing distance between periodic images of the projectile's path, rather than increasing supercell size alone.
Finally, we find that finite-size errors primarily influence free-electron contributions to stopping power, enabling convergence without the high computational costs incurred by explicitly modeling core electrons in large supercells.

Overall, our combination of approaches facilitates systematic control and analysis of two important approximations in first-principles stopping power calculations: the choice of projectile trajectory and finite supercells.
These strategies will not only enhance the accuracy and efficiency of TDDFT stopping power calculations, but also prove valuable for higher levels of theory~\cite{balzer:2016,moldabekov:2020} and quantum simulation algorithms on quantum computers \cite{babbush:2023} as their viabilities improve.

\section{Methods}
\label{sec:methods}

The real-time TDDFT calculations were performed with an in-house extension \cite{baczewski:2014,baczewski:2016,magyar:2016} of the Vienna \emph{ab initio} simulation package (\textsc{VASP}) \cite{kresse:1996a,kresse:1996b,kresse:1999}.
Ground-state orbitals from density functional theory with a Fermi smearing of \SI{100}{\kelvin} served as the initial condition for solving the time-dependent Kohn-Sham (KS) equations.
Plane-wave cutoff energies of \SI{750}{\electronvolt} achieved sufficiently converged results, and large supercells allowed reciprocal space sampling using the $\Gamma$ point only.
Sec.\ \ref{sec:sm:convergence} in the SI describes convergence of these parameters in more detail.
The adiabatic local density approximation \cite{zangwill:1980,zangwill:1981} was used for exchange and correlation (XC) after verifying accuracy relative to adiabatic versions of PBE \cite{perdew:1996} and SCAN \cite{sun:2015} (see \mbox{Sec.\ \ref{sec:sm:xc}} in the SI).

The electron-ion interaction was treated with the projector augmented-wave (PAW) method~\cite{blochl:1994}, explicitly including 3 or 11 valence electrons per aluminum atom.
The two aluminum pseudopotentials allowed access to different contributions to the stopping power, with 3-electron calculations isolating the response of free electrons, 11-electron calculations additionally including core contributions, and the difference between 11-electron and 3-electron results isolating core contributions.
Sec.\ \ref{sec:sm:pps} in the SI further discusses the influence of the pseudopotential approximation beyond the number of electrons explicitly modeled. 

The TDDFT simulations held ionic velocities fixed while propagating the KS orbitals according to the Crank-Nicolson algorithm.
For protons with velocities of 1.5 at.\,u.\ or more, the time step was chosen to scale inversely with proton velocity such that the proton traverses about \SI{0.02}{\angstrom} within each time step.
For slower protons, smaller time steps of 0.3\,--\,0.4 as were needed to achieve converged results (see Sec.\ \ref{sec:sm:convergence} of SI for more details).

The ultimate stopping powers were computed from the time-dependent force on the proton, including Hellmann-Feynman and Pulay terms, but not the fully energy-conserving form derived in Ref.\ \onlinecite{ojanpera:2012}, which is expected to be insignificant over the few-fs time scales simulated in this work.
Data during the first \mbox{4 \AA}\ of the proton's motion were excluded from the analysis to allow for dynamical ionization of the suddenly accelerated proton.
The time-dependent force was integrated along the proton's path to obtain the stopping work, or cumulative energy deposited by the proton into the electronic system.
The slope of the least-squares linear fit of the stopping work then produced the average stopping power.
In Sec.\ \ref{sec:sm:extraction} of the SI, we compare different procedures for extracting an average stopping power from TDDFT data and show that the methodology described here converges to within 2\% after the proton traverses \SI{75}{\angstrom}, whereas other procedures can be much more sensitive to the precise endpoint of the analysis and take longer to converge.

\section{Data Availability}
The datasets computed and analyzed during the current study are available from the corresponding author upon reasonable request.

\begin{acknowledgements}
We thank Andr\'e Schleife, Cheng-Wei Lee, Susan Atlas, Alexandra Olmstead, and Alexander White for helpful technical discussions.
We are also grateful to Joel Stevenson for technical support, Raymond Clay for feedback on the manuscript, and Heath Hanshaw for pre-publication review.
AK, ADB, and SBH were partially supported by the US Department of Energy Science Campaign 1.
SBH and TH were partially supported by the US Department of Energy, Office of Science Early Career Research Program, Office of Fusion Energy Sciences under Grant No.\ FWP-14-017426.
All authors were partially supported by Sandia National Laboratories' Laboratory Directed Research and Development (LDRD) Project No.\ 218456.

This article has been co-authored by employees of National Technology \& Engineering Solutions of Sandia, LLC under Contract No. DE-NA0003525 with the U.S. Department of Energy (DOE). The authors own all right, title and interest in and to the article and are solely responsible for its contents. The United States Government retains and the publisher, by accepting the article for publication, acknowledges that the United States Government retains a non-exclusive, paid-up, irrevocable, world-wide license to publish or reproduce the published form of this article or allow others to do so, for United States Government purposes. The DOE will provide public access to these results of federally sponsored research in accordance with the DOE Public Access Plan \url{https://www.energy.gov/downloads/doe-public-access-plan}.
\end{acknowledgements}

\bibliography{main.bib}

\clearpage
\onecolumngrid
\begin{center}
\textbf{\large Supplementary Information}
\vspace{0.1in}
\end{center}

\setcounter{page}{1}
\setcounter{section}{0}
\setcounter{figure}{0}
\setcounter{equation}{0}
\renewcommand{\thesection}{S\arabic{section}}
\renewcommand{\theHsection}{S\thesection}
\renewcommand{\thefigure}{S\arabic{figure}}
\renewcommand{\theHfigure}{S\thefigure}
\renewcommand{\theequation}{S\arabic{equation}}

\twocolumngrid

\section{Stopping power extraction}
\label{sec:sm:extraction}

An average stopping power may be extracted from the time-dependent data computed within TDDFT through various approaches.
One way is to time-average the force acting against the projectile's motion,
\begin{equation}
    S_\mathrm{avg} = -\frac{1}{T-t_0}\int_{t_0}^T \mathbf{F}(t)\cdot\hat{\mathbf{v}} \; dt,
    \label{eq:sm:avgf}
\end{equation}
where $\mathbf{F}(t)$ is the time-dependent force on the projectile, $\hat{\mathbf{v}}$ is the direction of the projectile's velocity, $T$ is the total simulation time, and $t_0>0$ typically excludes a transient regime at the beginning of the simulation.
The integral over time in Eq.\ \eqref{eq:sm:avgf} can be equivalently written as an integral over distance: 
\begin{align}
    S_\mathrm{avg} =& -\frac{1}{x_T - x_0}\int_{x_0}^{x_T} \mathbf{F}(x)\cdot\hat{\mathbf{v}} \; dx \\
    =& \frac{W(x_T) - W(x_0)}{x_T - x_0},
    \label{eq:sm:savg}
\end{align}
where $x$ is the distance traveled by the projectile, \mbox{$x_0=vt_0$}, $x_T=vT$, and 
\begin{equation}
    W(x) = -\int \mathbf{F}(x)\cdot\hat{\mathbf{v}} \; dx
    \label{eq:sm:work}
\end{equation}
is the so-called stopping work or the energy deposited by the projectile into the electronic system.

\begin{figure}
    \centering
    \includegraphics{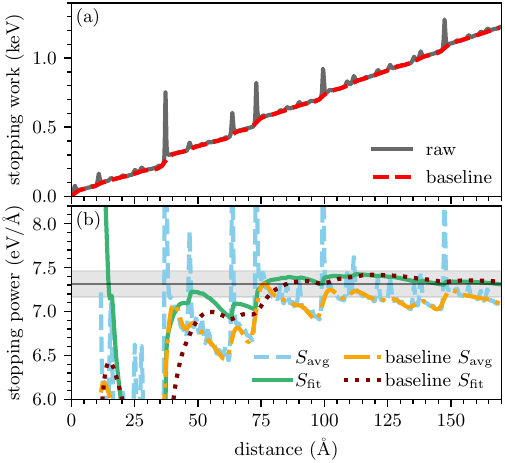}
    \caption{
    (a) Stopping work and (b) average stopping power computed according to different post-processing methods for a proton with 4 at.\,u.\ of velocity traversing an off-channeling trajectory in aluminum.
    The stopping work is evaluated according to Eq.\ \eqref{eq:sm:work}, while the stopping power is taken from either a two-point average (Eq.\ \eqref{eq:sm:savg}) or a least-squares linear fit (Eq.\ \eqref{eq:sm:sfit}) as a function of $x_T$ with $x_0=$\SI{4}{\angstrom} fixed.
    Results where sharp features in the stopping work were first smoothed through baseline fitting with asymmetric least squares \cite{eilers:2003} are also shown for comparison.
    Gray shading indicates values within 2\% of the final stopping power obtained with the least-squares fit.
    }
    \label{fig:sm:extraction}
\end{figure}

We see from Eq.\ \ref{eq:sm:savg} that $S_\mathrm{avg}$ amounts to a two-point approximation of the average energy deposition rate.
Therefore, the average stopping power extracted in this fashion can be sensitive to the precise choice of $x_0$ and $x_T$, particularly when core electron excitations during close collisions with host nuclei lead to sharp features in $W(x)$ (see Fig.\ \ref{fig:sm:extraction}).
A scheme to eliminate these sharp features such as the baseline method \cite{eilers:2003} suggested by Ref.\ \onlinecite{yost:2016} reduces sensitivity to $x_0$ and $x_T$, but $S_\mathrm{avg}$ still converges rather slowly with increasing $x_T$.

Instead of simply averaging the time-dependent force, one may perform a linear fit 
\begin{equation}
    W(x) \approx S_{\mathrm{fit}} \, x + W_0 \quad \mathrm{for} \quad x_0 \leq x \leq x_T
    \label{eq:sm:sfit}
\end{equation}
and take the resulting slope $S_\mathrm{fit}$ as the average stopping power.
Although the two methods should eventually converge to the same value,
we find that this latter method produces average stopping powers that evolve more smoothly as the simulation proceeds, i.e., as $x_T$ increases, and approach a constant average stopping power earlier (see Fig.\ \ref{fig:sm:extraction}).
Therefore, the linear fitting procedure allows more precise determination of stopping powers with shorter simulation times and thus smaller supercells.
We similarly find that linear fitting reduces sensitivity to $x_0$, as keeping $x_T\geq$ \SI{80}{\angstrom} fixed and varying $x_0$ from 0 to \SI{10}{\angstrom} produces stopping powers within 2\% of the converged value shown in Fig.\ \ref{fig:sm:extraction}.
Throughout the main text, we employ least-squares linear fits with $x_0=4$\,\AA\ and $x_T\approx 80$\,\AA.
Since the baseline filtering of the stopping work barely alters the fitted stopping power for $x_T>80$\,\AA, we do not bother with this complication in the main text.

\section{Sampling nearest-neighbor distributions}
\label{sec:sm:dists}

\begin{figure}[h]
    \centering
    \includegraphics{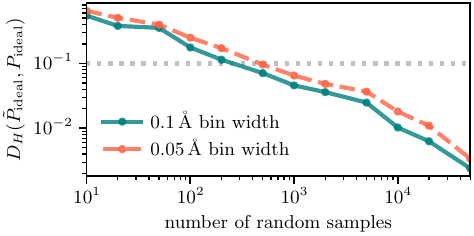}
    \caption{Convergence of the ideal nearest-neighbor distribution as the number of randomly sampled points increases.
    Hellinger distances (see Eq.\ \eqref{eq:dh} in the main text) between an ideal distribution $P_\mathrm{ideal}(\delta_{NN})$ sampled using 60,000 points and a more approximate ideal distribution $\tilde{P}_\mathrm{ideal}(\delta_{NN})$ sampled using a variable number of points are shown for two different histogram bin widths.
    The gray dotted line indicates the 0.1 $D_H$ threshold identified in the main text for ``good" off-channeling trajectories.
    }
    \label{fig:sm:Pideal_conv}
\end{figure}

In the main text, the ideal nearest-neighbor distribution $P_{\mathrm{ideal}}(\delta_{NN})$ is sampled using 50,000 random points and histogrammed using a \SI{0.1}{\angstrom} bin width.
The Hellinger distance between this distribution and an ideal distribution sampled using 60,000 random points is only 0.002, well below the 0.1 threshold established in the main text for selecting representative trajectories (see \mbox{Fig.\ \ref{fig:sm:Pideal_conv}}).
Thus, the ideal nearest-neighbor distribution used throughout the main text is very well-converged for our purposes in this work.
We note that the number of samples needed depends on the bin width, with smaller bins requiring more points.

\begin{figure}
    \centering
    \includegraphics{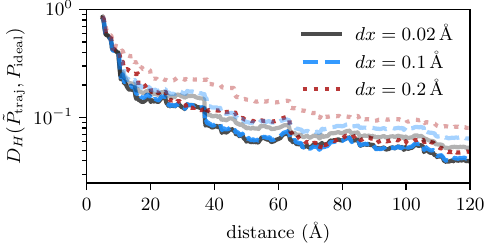}
    \caption{Convergence of the nearest-neighbor distributions $\tilde{P}_\mathrm{traj}$ achieved by a given projectile trajectory as the sampling step size $dx$ decreases and the number of sampled points increases.
    Hellinger distances from the ideal distribution are compared for histogram bin widths of \SI{0.1}{\angstrom} (dark curves) and \SI{0.05}{\angstrom} (light curves).
    }
    \label{fig:sm:Ptraj_conv}
\end{figure}

We sample the nearest-neighbor distribution $P_{\mathrm{traj}}(\delta_{NN})$ achieved by a particular trajectory by taking discrete steps of about \SI{0.02}{\angstrom} along that trajectory.
This step length matches the typical distance traveled by the projectile in each time step of the ultimate TDDFT simulations, and it produces around 10 times fewer samples than the number used to construct $P_{\mathrm{ideal}}(\delta_{NN})$.
However, sampling more points along the trajectory without increasing its total length does not significantly alter $P_{\mathrm{traj}}$ because the nearest-neighbor distance evolves continuously along the projectile's path and thus closely-spaced samples are highly correlated.
In fact, the step size only begins to noticeably affect computed $D_H$ values when it exceeds the histogram bin width of \SI{0.1}{\angstrom} (see \mbox{Fig.\ \ref{fig:sm:Ptraj_conv}}).

\section{Comparison between Hellinger distance and overlap index}
\label{sec:sm:gu_comparison}
As described in Section \ref{sec:traj} of the main text, we evaluate the extent to which individual trajectories representatively sample the simulation cell using the Hellinger distance $D_H$ (see Eq.\ \eqref{eq:dh} in the main text).
Gu et al.\ \cite{gu:2020} took an alternative approach based on the overlap index
\begin{equation}
    O = 1-\frac{1}{2}\int \left|P_\mathrm{traj}(\delta_{NN}) - P_\mathrm{ideal}(\delta_{NN})\right| \; d\delta_{NN},
    \label{eq:sm:overlap}
\end{equation}
which quantifies the overlapping area under $P_\mathrm{traj}$ and $P_\mathrm{ideal}$.
Minimizing $D_H$, as proposed in this work, and maximizing $O$, as done in \mbox{Ref.\ \onlinecite{gu:2020}}, represent different ways to optimize trajectories with the same underlying principle that the nearest-neighbor distribution sampled along the trajectory should resemble the ideal or reference distribution.

\begin{figure}
    \centering
    \includegraphics{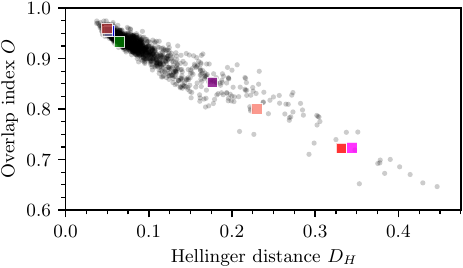}
    \caption{Comparison between the Hellinger distance $D_H$ used in this work (see Eq.\ \eqref{eq:dh} in the main text) and the overlap index $O$ used by Ref.\ \onlinecite{gu:2020} (see Eq.\ \eqref{eq:sm:overlap}) to evaluate how closely the nearest-neighbor distribution sampled by individual trajectories matches the ideal nearest-neighbor distribution.
    Translucent gray points represent the same 1440 random trajectories analyzed in Fig.\ \ref{fig:traj_random} of the main text, while colored squares represent trajectories analyzed in Fig.\ \ref{fig:traj_metric} of the main text.
    }
    \label{fig:sm:dh_vs_o}
\end{figure}

The $D_H$ method offers the advantages of a mathematical metric: it satisfies the triangle inequality and thus allows rigorous comparisons among different trajectories in addition to evaluating the similarity of $P_\mathrm{traj}$ and $P_\mathrm{ideal}$.
Although $O$ does not have the mathematical properties of a metric, we note that $1-O$ does.
Furthermore, we find that $O$ and $D_H$ are strongly correlated (see \mbox{Fig.\ \ref{fig:sm:dh_vs_o}}), and we expect that either option will result in suitably representative trajectory selections.
However, detailed comparisons between similar quality trajectories could depend on the metric used, since $D_H$ and $O$ can lead to opposite conclusions for which of two trajectories matches the ideal distribution more closely.

\section{Convergence of TDDFT simulations}
\label{sec:sm:convergence}

To establish computational parameters that achieve converged stopping powers, we performed various tests for off-channeling protons with 4 at.\,u.\ of velocity traversing aluminum.
Since this velocity lies above the Bragg peak (see Fig.\ \ref{fig:Scold} in the main text), these tests probe convergence of contributions from both free and core electrons in aluminum.
Smaller supercells containing 64, 96, or 108 aluminum atoms were used to reduce computational costs.

\begin{figure}
    \centering
    \includegraphics[width=\columnwidth]{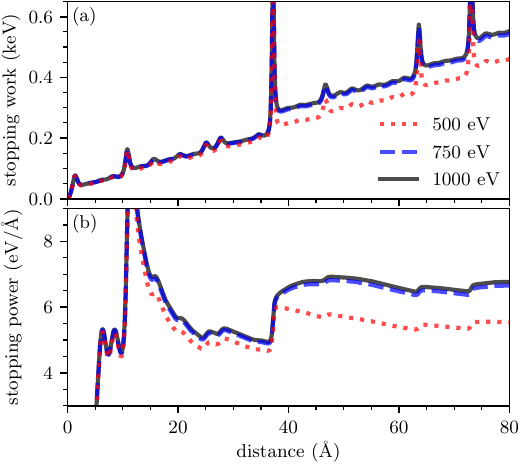}
    \caption{(a) Stopping work and (b) fitted stopping power computed for a proton with 4 at.\,u.\ of velocity traversing an off-channeling trajectory in aluminum (64-atom cell) with different plane-wave cutoff energies.}
    \label{fig:sm:ecut}
\end{figure}

First, we considered the influence of plane-wave cutoff energy and found that results computed with the \SI{750}{\electronvolt} cutoff used throughout the main text are converged within 1.7\% relative to a \SI{1000}{\electronvolt} cutoff.
Throughout most of the simulation, the instantaneous forces on the proton are extremely well-converged even with a lower cutoff of \SI{500}{\electronvolt}, which suffices to capture the free-electron contribution to stopping power given by the slope of the nearly linear portions of the stopping work plotted in \mbox{Fig.\ \ref{fig:sm:ecut}a}.
However, converging core-electron contributions requires higher cutoff energies: a \SI{500}{\electronvolt} cutoff significantly underestimates energy transferred to core electrons during close collisions such as the one occurring at around \SI{37}{\angstrom} in Fig.\ \ref{fig:sm:ecut}, resulting in an 18\% underestimation of the ultimate stopping power.

These findings illustrate the need for careful scrutiny of core-electron contributions during convergence testing of electronic stopping calculations.
Details of the pseudopotentials naturally influence the convergence behavior of core excitations.
For instance, under the harder, norm-conserving pseudopotentials used within Qb@ll, stopping powers computed with \SI{750}{\electronvolt} and \SI{1000}{\electronvolt} cutoff energies differ by 3\%, somewhat more than in the VASP tests described above.
Therefore, in Qb@ll calculations we used a higher cutoff of \SI{1000}{\electronvolt} to obtain results converged within 1\% relative to a \SI{1250}{\electronvolt} cutoff.

Next, we find that reciprocal-space sampling has a negligible effect on the accuracy of our results.
For a 108-atom aluminum supercell, the difference between stopping power computed with the $\Gamma$ point only and with a $\Gamma$-centered $2\times 2\times 2$ k-point grid is only 0.2\%.
The influence of k-point sampling should be even more minuscule for the 256-atom supercell used in production calculations.

Finally, we considered convergence with respect to numerical time step.
For protons with at least 1.5\,at.\,u. of velocity, we scaled the time step inversely with velocity such that the projectile traversed about \SI{0.02}{\angstrom} in each step.
For slower protons, smaller steps of 0.3\,--\,\SI{0.4}{\atto\second} were needed to adequately resolve the electron dynamics.
These choices produced stopping powers converged within 1.6\% or less relative to tests where the time step was halved.

\section{Pseudopotentials}
\label{sec:sm:pps}
Pseudopotentials are widely known to influence stopping power calculations through their inclusion or exclusion of core excitations depending on the number of valence electrons explicitly modeled.
We exploit this fact in the main text to isolate stopping power contributions from core and free electrons, allowing a cost-reducing scheme for converging finite-size effects.
However, we also find a more subtle dependence on the type and details of the pseudopotentials even with a fixed number of valence electrons.

\begin{figure}
    \centering
    \includegraphics{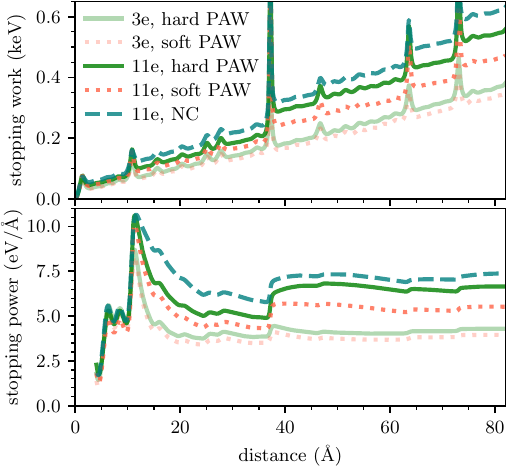}
    \caption{(a) Stopping work and (b) fitted stopping power computed for a proton with 4 at.\,u.\ of velocity traversing an off-channeling trajectory in aluminum using different pseudopotentials for both the projectile and host ions.
    Light (dark) curves explicitly treat 3 (11) electrons per aluminum atom in a 256-atom (64-atom) cell.
    Solid green (dotted orange) curves were computed using PAW within VASP and a hard (soft) hydrogen pseudopotential representing the projectile.
    The dashed teal curve was computed using norm-conserving pseudopotentials within Qbox/Qb@ll.
    }
    \label{fig:sm:pp}
\end{figure}

Fig.\ \ref{fig:sm:pp} compares results computed using PAW \cite{blochl:1994} within VASP \cite{baczewski:2016,magyar:2016} and norm-conserving pseudopotentials \cite{vanderbilt:1985,troullier:1991} within Qbox/Qb@ll \cite{schleife:2012,draeger:2017}.
In addition to aluminum pseudopotentials with 3 or 11 valence electrons per atom, we also tested two hydrogen pseudopotentials within the VASP calculations: the standard one among those distributed with the package (labeled ``H") and a harder one (labeled ``H\_h").
The Al and H pseudopotentials used in the Qb@ll calculation both have smaller core radii than the VASP pseudopotentials, though the augmentation charge within the PAW method complicates the comparison.
The plane-wave cutoff energy and time step were separately converged for the Qb@ll calculation, where the harder pseudopotentials required a higher cutoff of \SI{1000}{\electronvolt} and a time step of about \SI{1}{\atto\second} sufficed.
Other parameters, including supercell size, projectile trajectory, exchange-correlation functional, and Fermi smearing remained consistent across the VASP and Qb@ll calculations so as to isolate the influence of the pseudopotentials.

We find significant differences among stopping powers computed with the different pseudopotentials even when the number of valence electrons is held fixed.
Within VASP, the standard, softer H pseudopotential leads to stopping powers 0.3 (0.8) \SI{}{\electronvolt\per\angstrom} or 8\% (13\%) below those computed with the harder H pseudopotential when explicitly treating 3 (11) electrons per Al atom.
The even harder norm-conserving pseudopotentials used in the Qb@ll calculation further increase the stopping power by \SI{0.6}{\electronvolt\per\angstrom} or 10\% relative to the hard VASP calculation.
Dynamics during close collisions (e.g., at around 11 and \SI{37}{\angstrom} in Fig.\ \ref{fig:sm:pp}) contribute significantly to these discrepancies, as softer pseudopotentials tend to underestimate energy transferred during these events.
These findings indicate the need for future work to further characterize the influence of the pseudopotential approximation on stopping power calculations and develop computationally efficient approaches for mitigating this relatively large source of uncertainty in computed results.

\section{Exchange and Correlation}
\label{sec:sm:xc}

As with all DFT-based methods, the choice of exchange-correlation (XC) functional remains a central approximation in our work with few avenues for improving accuracy.
We assessed the sensitivity of our results to the XC functional by comparing adiabatic LDA \cite{zangwill:1980,zangwill:1981}, PBE \cite{perdew:1996}, and SCAN \cite{sun:2015} for the case of an off-channeling proton moving through aluminum with 4 at.\,u.\ of velocity.
When using the same pseudopotentials and otherwise identical parameters, stopping powers computed with these XC functionals differ among each other by 0.5\% or less.

Notably, these variations are much smaller than an earlier report of up to 18\% differences between stopping powers computed with LDA and PBE for SiC \cite{yost:2016}, a semiconductor that can be expected to exhibit higher sensitivity to XC effects.
We conclude that for metals, the influence of the XC functional appears very minor compared to other factors typically limiting accuracy of stopping power calculations, namely finite-size effects, trajectory dependence, and the pseudopotential approximation.
However, we note that non-adiabatic XC effects on electronic stopping power remain unexplored.

\end{document}